\def\year{2016}
\documentclass[letterpaper]{article}
\pdfoutput=1
\usepackage{aaai16}
\usepackage{times}
\usepackage{helvet}
\usepackage{courier}
\usepackage{graphicx}
\usepackage{amsmath}

\newcommand{\specialcell}[2][c]{%
  \begin{tabular}[#1]{@{}c@{}}#2\end{tabular}}
\graphicspath{{images/}}

\title{The Status Gradient of Trends in Social Media}
\author{Rahmtin Rotabi \\
Department of Computer Science \\
Cornell University \\
Ithaca, NY, 14853 \\
\texttt{rahmtin@cs.cornell.edu}
\And Jon Kleinberg\\
Department of Computer Science \\
Cornell University \\
Ithaca, NY, 14853\\
\texttt{kleinber@cs.cornell.edu}}

\begin{document}
\maketitle
\begin{abstract}
An active line of research has studied the detection and
representation of trends in social media content. There is still
relatively little understanding, however, of methods to characterize
the early adopters of these trends: who picks up on these trends
at different points in time, and what is
their role in the system?  We develop a framework for analyzing
the population of users who participate in trending topics over
the course of these topics' lifecycles.  Central to our analysis is
the notion of a {\em status gradient}, describing how users of
different activity levels adopt a trend at different points in time.
Across multiple datasets, we find that this 
methodology reveals key differences in the nature of the early
adopters in different domains.

\end{abstract}	

\section{Introduction}
\newcommand{\xhdr}[1]{\paragraph*{\bf #1.}}
\newcommand{\xhdrb}[1]{\paragraph*{\bf #1}}

An important aspect of the everyday experience on large on-line
platforms is the emergence and spread of new activities and behaviors,
including resharing of content, participation in new topics,
and adoption of new features.
These activities are described by various terms --- as
{\em trends} in the topic detection and social media literatures, and 
{\em innovations} by sociologists working on the diffusion of
new behaviors \cite{rogers-diffusion}.

An active line of recent research has used rich Web datasets to study the
properties of such trends in on-line settings, and how they develop over time
(e.g. \cite{adar-blogspace,gruhl-blogspace,liben-nowell-pnas08,leskovec-ec06j,dow2013anatomy,goel2012structure,aral09influence,backstrom-kdd06,wu-who-says-what-www11}).
The analyses performed in this style have extensively investigated the
temporal aspects of trends, including patterns that
accompany bursts of on-line activity
\cite{kleinberg-kdd02,kumar-bursty-blogspace,Crane2008,yang2011patterns},
and the network dynamics of their spread at both local levels
\cite{backstrom-kdd06,leskovec-ec06j} and global levels
\cite{liben-nowell-pnas08,dow2013anatomy,goel2012structure}.

An issue that has received less exploration using these types of datasets
is the set of
distinguishing characteristics of the participants themselves --- those
who take part in a trend in an on-line domain.
This has long been a central question for sociologists working in 
diffusion more broadly: who adopts
new behaviors, and how do early adopters differ from later ones
\cite{rogers-diffusion}?

\xhdrb{Key question: Who adopts new behaviors, and when do they adopt them?}
When empirical studies of trends and innovations in off-line domains seek
to characterize the adopters of new behaviors,
the following crucial dichotomy emerges:
is the trend proceeding from the ``outside in,'' starting with
peripheral or marginal members of the community and being adopted
by the high-status central members; or is the innovation proceeding 
from the ``inside out,'' starting with the elites and moving
to the periphery \cite{abrahamson-diffusion-simulation,becker-socio-location,crossan-org-innovation,daft-dual-core-innov,pampel-status-gradient-smoking}?

Note that this question can be framed at either a broader population
level or a more detailed network structural level.
We pursue the broader population-level
framing here, in which it is relevant to any
distinction between elite and more peripheral members of a community,
and not necessarily tied to measures based on network structure.

There are compelling arguments for the role of both the elites
and the periphery in the progress of a trend.
Some of the foundational work on adopter characteristics established 
that early adopters have significantly 
higher socioeconomic status in aggregate than arbitrary members
of the population \cite{deutschmann-adopter-characteristics,rogers-adopter-characteristics}; elites also play a crucial role --- as likely 
{\em opinion leaders} ---in the two-step flow theory of media influence
\cite{katz-personal-influence}.
On the other hand, a parallel line of work has argued for the important
role of peripheral members of the community in the emergence of
innovations; Simmel's notion of ``the stranger'' who brings ideas
from outside the mainstream captures this notion 
\cite{simmel-sociology}, as does the theory of {\em change agents}
\cite{mclaughlin-change-agent-revisited,valente-network-interventions}
and the power of individuals who span {\em structural holes},
often from the periphery of a group
\cite{burt-good-ideas,krackhardt-org-viscosity}.

This question of how a trend flows through a population ---
whether from high-status individual to lower-status ones, or vice versa --- 
is a deep issue at the heart of diffusion processes.
It is therefore natural to ask how it is reflected in the adoption
of trends in on-line settings.
The interesting fact, however, is that there is no existing general framework
or family of measures that can be applied to user activity in an
on-line domain to characterize trends according to whether they
are proceeding from elites outward or peripheral members inward.
In contrast to the extensive definitions and measures that
have been developed to characterize temporal and network properties
of on-line diffusion, this progress of adoption along dimensions of status
is an issue that to a large
extent has remained computationally unformulated.

\xhdr{The present work: Formulating the status gradient of a trend}
In this paper, we define a formalism that we term the 
{\em status gradient}, which aims to take a first step
toward characterizing how the adopter
population of a trend changes over time with respect to their
status in the community.
Our goal in defining the status gradient is that it should be easy to 
adapt to data from different domains, and it should
admit a natural interpretation
for comparing the behavior of trends across these domains.

We start from the premise that the computation
of a status gradient for a trend should produce a time series 
showing how the status of adopters in the community 
evolves over the life cycle of the trend.
To make this concrete, we need
(i) a way of assessing the status of community members, and
(ii) a way of identifying trends.

While our methods can adapt to any way of 
defining (i) and (ii), for purposes of the present paper we
operationalize them in a simple, concrete way as follows.
Since our focus in the present paper is on settings where
the output of the community is textual, we will think abstractly of each user
as producing a sequence of posts, and the candidate trends
as corresponding to words in these posts.
(The adaptation to more complex definitions of status and trends 
would fit naturally within our framework as well.)
\begin{itemize}
\item We will use the activity level of each user as a simplified
proxy for their status: users who produce more content are in 
general more visible and more actively engaged in the community, and 
hence we can take this activity as a simple form of 
status.\footnote{In the datasets with a non-trivial presence
of high-activity spammers, we employ heuristics to remove such users,
so that this pathological form of high activity is kept out of our
analysis.}
The current activity level of a user 
at a time $t$ is the total number of posts they have produced up until $t$,
and their final activity level is the total number of posts they have 
produced overall.  
\item 
We use a burst-detection approach for identifying trending words in posts
\cite{kleinberg-kdd02}; thus, for a given trending word $w$, we
have a time $\beta_w$ when it entered its {\em burst state} of elevated activity.
When thinking about a trending word $w$, we will generally 
work with ``relative time'' in which $\beta_w$ corresponds to time $0$.
\end{itemize}

We could try to define the status gradient simply in terms of 
the average activity level (our proxy for status) of the users
who adopt a trend at each point in time.
But this would miss a crucial point: high-activity users 
are already overrepresented in trends simply because they are
overrepresented in {\em all} of a site's activities.
This is, in a sense, a consequence of what it means to be high-activity.
And this subtlety is arguably part of the reason why a useful
definition of something like the status gradient has been elusive.

Our approach takes this issue into account.
We provide precise definitions in the following section, but
roughly speaking we say that the status gradient for a trending word $w$
is a function $f_w$ of time, where $f_w(t)$ measures the extent to
which high-activity users are overrepresented or underrepresented in
the use of $w$, {\em relative} to the baseline distribution of activity levels
in the use of random words. 
The point is that since high-activity users are expected to be
heavily represented in usage of both $w$ and of ``typical'' words,
the status gradient is really emerging from the difference between
these two.

\section{Overview of Approach and Summary of Results} \label{sec:overview}
We apply our method to a range of on-line datasets, including 
Amazon reviews from several large product categories \cite{mcauley-amazon}, 
Reddit posts and comments from several active sub-communities \cite{tan-reddit},
posts from two beer-reviewing communities \cite{danescu-no-country-www13},
and paper titles from DBLP and Arxiv.

We begin with a self-contained description of the status gradient
we compute, before discussing the detailed implementation and results
in subsequent sections.
Recall that for purposes of our exposition here,
we have an on-line community containing posts by users;
each user's activity level is the number of posts he or she has produced;
and a trending word $w$ is a word that appears in a subset of the posts
and has a burst starting at a time $\beta_w$.

Perhaps the simplest attempt to define a status gradient would be
via the following function of time.
First, abusing terminology slightly, we define the activity level
{\em of a post} to be the activity level of the post's author.
Now, let $P_w(t)$ be the set of all posts at time $t + \beta_w$ containing $w$,
and let $g_w(t)$ be the median activity level of the posts in $P_w(t)$.

Such a function $g_w$
would allow us to determine whether the median activity level
of users of the trending word $w$ is increasing or decreasing with time,
but it would not allow us to make statements about whether
this median activity level at a given time $t + \beta_w$ is high or low
viewed as an isolated quantity in itself.
To make this latter kind of statement, we need a baseline for comparison,
and that could be provided most simply by comparing $g_w(t)$ to
the median activity level $g^*$ of the set of {\em all} posts in the community.

The quantity $g^*$ has an important meaning: half of all posts are
written by users of activity level above $g^*$, and half are written
by users of activity level below $g^*$.
Thus if $g_w(t) < g^*$, it means that the users of activity
level at most $g_w(t)$ are producing half the occurrences of $w$
at time $t + \beta_w$, but globally are producing less than half the posts
in the community overall.
In other words, the trending word $w$ at time $t$ is being overproduced
by low-activity users and underproduced by high-activity users;
it is being adopted mainly by the periphery of the community.
The opposite holds true if $g_w(t) > g^*$.
Note how this comparison to $g^*$ allows us to make absolute
statements about the activity level of users of $w$ at time $t + \beta_w$
without reference to the activity at other times.

This then suggests how to define the status gradient function
$f_w(t)$ that we actually use, as a normalized version of $g_w(t)$.
To do so, we first define the distribution of activity levels
$H : [0,\infty) \rightarrow [0,1]$
so that $H(x)$ is the fraction of {\em all} posts
whose activity level is at most $x$.
We then define
$$f_w(t) = H(g_w(t)).$$
This is the natural general formulation of our observations in
the previous paragraph: 
the users of activity
level at most $g_w(t)$ are producing half the occurrences of $w$
at time $t + \beta_w$, but globally are producing an $f_w(t)$ fraction
of the posts in the community overall.
When $f_w(t)$ is small (and in particular below $1/2$), it means
that half the occurrences of $w$ at time $t + \beta_w$ are being produced
by a relatively small slice of low-activity users, so the trend
is being adopted mainly by the periphery;
and again, the opposite holds when $f_w(t)$ is large.

Our proposal, then, is to consider $f_w(t)$ as a function of time.
Its relation to $1/2$ conveys whether the trend is being
overproduced by high-activity or low-activity members of the community,
and because it is monotonic in the more basic function $g_w(t)$,
its dynamics over time show how this effect changes over the
life cycle of the trend $w$.

\xhdr{Summary of Results}
We find recurring patterns in the status gradients that reflect
aspects of the underlying domains.  
First, 
for essentially all the datasets,
the status gradient indicates that high-activity users are 
overrepresented in their adoption of trends (even relative
to their high base rate of activity), suggesting their role
in the development of trends.

We find interesting behavior in the status gradient right around time $0$,
the point at which the burst characterizing the trend begins.
At time $0$, the status gradient for most of the sites we study
exhibits a sharp drop, 
reflecting an influx of lower-activity users as the trend
first becomes prominent.
This is a natural dynamic; however, it is not the whole picture.
Rather, for datasets where we can identify a distinction between
{\em consumers} of content (the users creating posts on the site)
and {\em producers} of content (the entities generating the primary
material that is the subject of the posts), we generally find
a sharp drop in the activity level of consumers at time $0$,
but {\em not} in the activity level of producers.
Indeed, for some of our largest datasets, the activity levels of
the two populations move inversely at time $0$, with the activity
level of consumers falling as the activity level of producers rises.
This suggests a structure that is natural in retrospect but
difficult to discern without
the status gradient: in aggregate, the onset of a burst is
characterized by producers of rising status moving in to provide content
to consumers of falling status.

We now provide more details about the methods and the datasets
where we evaluate them, followed by the results we obtain.

\section{Data Description}

Throughout this paper, we will study multiple on-line
communities gathered from different sources. The study uses the
three biggest communities on Amazon.com, several of the largest 
sub-reddits from reddit.com,  two large beer-reviewing communities
that have been the subject of prior research, and the set of
all papers on DBLP and Arxiv (using only the title of each).
\begin{itemize}
\item Amazon.com, in addition to allowing users to purchase items,
hosts a rich set of reviews; these are the textual posts that
we use as a source of trends.
We take all the reviews written before December
2013 for the top 3 departments: TV and Movies, Music, and Books
\cite{mcauley-amazon}. 
\item Reddit is one of the most active community-driven
platforms, allowing users to post questions, ideas and comments.
Reddit is organized into thousands of categories called sub-reddits;
we study 5 of the biggest text-based sub-reddits. Our Reddit data
includes all the Reddit posts and comments posted
before January 2014 \cite{tan-reddit}. 
Reddit contains a lot of content generated by robots and spammers;
heuristics were used to remove this content from the dataset.
\item The two on-line beer communities include reviews
of beers from 2001 to 2011. Users on these two platforms describe a
beer using a mixture of well-known and newly-adopted
adjectives \cite{danescu-no-country-www13}. 
\item DBLP is a website with
bibliographic data for published papers in the computer science
community.  For this study we only use the title of the publications. 
\item Arxiv is a repository of on-line preprints of scientific papers 
in physics, mathematics, computer science, and an expanding set
of other scientific fields.
As with DBLP, we use the titles of the papers uploaded on Arxiv 
for our analysis, restricting to papers before November 2015.
We study both the set of all Arxiv papers (denoted \textit{Arxiv All}), 
as well as subsets corresponding to well-defined sub-fields.
Two that we focus on in particular
are the set of all statistics and computer science
categories, denoted \textit{Arxiv stat-cs}, and astrophysics ---
denoted \textit{Arxiv astro-ph} --- as an
instance of a large sub-category of physics.
In this
study we only use papers that use \textit{\textbackslash author} and
\textit{\textbackslash title} for including their title and their
names.
\end{itemize}

More specific details
about these datasets can be found in Table \ref{Table:DataSpec}.

\begin{table}
\centering
\begin{tabular}{|c|c|c|} \hline
Dataset & Authors & Documents \\ \hline
Amazon Music & 971,186 & 11,726,645 \\ \hline
Amazon Movies and TV & 846,915 & 14,391,833  \\ \hline
Amazon Books & 1,715,479 & 23,625,228\\ \hline
Reddit music & 969,895 & 5,873,797 \\ \hline
Reddit movies & 930,893 & 1,0541,409\\ \hline
Reddit books & 392,000 & 2,575,104\\ \hline
Reddit worldnews & 1,196,638 & 16,091,492\\ \hline
Reddit gaming & 1,811,850 & 33,868,254 \\ \hline
Rate Beer & 29,265 & 2,854,842 \\ \hline
Beer Advocate & 343,285 & 2,908,790\\ \hline
DBLP & 1,510,698 & 2,781,522\\ \hline
Arxiv astro-ph & 83,983 & 167,580\\ \hline
Arxiv stats-cs & 63,128 & 71,131 \\ \hline
Arxiv All &  326,102 & 717,425  \\ \hline
\end{tabular}

\caption{Number of authors and documents in the studied datasets.}
\label{Table:DataSpec}
\end{table}

\section{Details of Methods}
In this section we describe our method for finding trends
and then how we use these to compute the status gradient.
We run this
method for each of these datasets separately so we can compare the
communities with each other. 
In each of these communities, users produce textual content, 
and so for unity of terminology we will refer to the textual output
in any of these domains (in the form of posts, comments, reviews,
and publication titles) as a set of {\em documents};
similarly, we will refer to the producers of any of this content
(posters, commenters, reviewers, researchers) as the {\em authors}.
For Amazon, Reddit, and the beer communities we use an approach
that is essentially identical across all the domains;
the DBLP and Arxiv datasets have a structure that necessitates some slight
differences that we will describe below.

\subsection{Finding Trends}

As discussed above, the trends we analyze are associated with
{\em word bursts} --- words that increase in usage in a well-defined way.
We compute word bursts using an underlying probabilistic
automaton as a generative model, following \cite{kleinberg-kdd02}.
These word bursts form the set of trends on which we then base
the computation of status gradients.

For each dataset (among Amazon, Reddit, and the beer communities),
and for each word $w$ in the dataset, let $\alpha_w$ denote the
fraction of documents in which it appears.
We define a two-state automaton that we imagine to probabilistically generate
the presence or absence of the word $w$ in each document.
In its ``low state'' $q_0$ the automaton generates the word with probability
$\alpha_w$, and in its ``high state'' $q_1$ it generates the word with
probability $c_D \alpha_w$ for a constant $c_D > 1$ that is uniform
for the given dataset $D$.
Finally, it transitions between the two states with probability $p$.
(In what follows we use $p = 0.1$, but other values give similar results.)

Now, for each word $w$, let
$f_{w,1}, f_{w,2}, \ldots, f_{w,n}$ be a sequence in which 
$f_{w,i}$ denotes the fraction of documents in week $i$ that contain $w$.
We compute the state sequence 
$S_{w,1}, S_{w,2}, \ldots, S_{w,n}$ (with each $S_{w,i} \in \{q_0,q_1\}$)
that maximizes the likelihood of observing the fractions
$f_{w,1}, f_{w,2}, \ldots, f_{w,n}$ when the automaton starts in $q_0$.
Intuitively, this provides us with a sequence of ``low rate'' and
``high rate'' time intervals that conform as well as possible to the
observed frequency of usage, taking into account (via the transition
probability $p$) that we do not expect extremely rapid transitions
back and forth between low and high rates.
Moreover, words that are used very frequently throughout the duration of the
dataset will tend to produce state sequences that stay in $q_0$,
since it is difficult for them to rise above their already high rate
of usage.

A {\em burst} is then a maximal sequence of states that are all equal to
$q_1$, and the beginning of this sequence corresponds to a point in time
at which $w$ can be viewed as ``trending.''
The {\em weight} of the burst is the difference in log-probabilities
between the state sequence that uses $q_1$ for the
interval of the burst and the sequence that stays in $q_0$.

To avoid certain pathologies in the trends we analyze, 
we put in a number of heuristic filters;
for completeness we describe these here.
First, since a word might produce several disjoint time intervals
in the automaton's high state, we focus only on the interval with 
highest weight.  For simplicity of phrasing,
we refer to this as {\em the} burst for the word.
(Other choices, such as focusing on the 
first or longest interval, produce similar results.)
Next, we take a number of steps to make sure we are studying bursty
words that have enough overall occurrences, and that exist 
for more than a narrow window of time.
The quantity $c_D$ defined above controls how much higher the rate of $q_1$
is relative to $q_0$; too high a value of $c_D$ tends to produce 
short, extremely high bursts that may have very few occurrences of the word.
We therefore choose the maximum $c_D$ subject to the property that
the median number of occurrences of words that enter the burst state
is at least 5000.
Further, we only consider word bursts of at least eight weeks in length,
and only for words that were used at least once every three months
for a year extending in either direction from the start of the burst.

With these steps in place, we take the top 500 bursty words sorted by
the weight of their burst interval, and we use these as the trending
words for building the status gradient.
With our heuristics in place, each of these words occurred at least 200 times.
For illustrative purposes, 
a list of top 5 words for each dataset is shown in Table \ref{TopWords}.

\begin{table*}[ht]
\centering
\scalebox{0.95}{
\begin{tabular}{|c|c|c|} \hline
Dataset & Words & Bigrams \\ \hline
Amazon Music & anger, metallica, coldplay, limp, kanye & \specialcell{st-anger, green-day, \\limp-bizkit, 50-cent, x-y} \\ \hline
Amazon Movies and TV & lohan, lindsay, sorcerers, towers, gladiator &	 \specialcell{mean-Girls, rings-trilogy, lindsay-lohan,\\ matrix-reloaded, two-towers}  \\ \hline
Amazon Books & kindle, vinci, bush, phoenix, potter & \specialcell{da-Vinci, john-kerry, harry-potter, \\ twilight-book, fellowship-(of the)-ring}\\ \hline
Reddit Music & daft, skrillex, hipster, radiohead, arcade & \specialcell{daft-punk, get-lucky, chance-rapper,\\ mumford-(\&)-sons, arctic-monkeys}\\ \hline
Reddit movies & batman, bane, superman, bond, django & \specialcell{pacific-rim, iron-man, man-(of)-steel,\\ guardians-(of the)-galaxy, dark-knight} \\ \hline
Reddit books & hunger, nook, borders, gatsby, twilight & \specialcell{hunger-games, shades-(of)-grey, gone-girl,\\ great-gatsby, skin-game}\\ \hline
Reddit worldnews & israel, hamas, isis, gaza, crimea & \specialcell{north-korea, chemical-weapons, human-shields,\\ iron-dome, civilian-casualties} \\	 \hline
Reddit gaming & gta, skyrim, portal, diablo, halo & \specialcell{gta-v, last-(of)-us, mass-effect,\\ bioshock-infinite, wii-u}  \\ \hline
Rate Beer & cigar, tropical, winter, kernel, farmstead & \specialcell{cigar-city, black-ipa, belgian-yeast,\\ cask-handpump, hop-front}\\ \hline
Beer Advocate & finger, tulip, pine, funk, roast & \specialcell{lacing-s, finger-head, moderate-carbonation,\\ poured-tulip, head-aroma}\\ \hline
DBLP & parallel, cloud,  social, database, objectoriented & --- \\ \hline
Arxiv astro-ph & chandra, spitzer, asca, kepler, xmmnewton & ---\\ \hline
Arxiv stats-cs & deep, channels, neural, capacity, convolutional & --- \\ \hline
Arxiv All & learning, chandra, xray, spitzer, bayesian & --- \\ \hline
\end{tabular}
}
\caption{The top 5 words and bigrams that our algorithm finds using the
burst detection method. Words in parenthesis are stop words that got removed by the algorithm.}
\label{TopWords}
\end{table*} 
\subsection{Computing the Status Gradient}

Now we describe the computation of the status gradient.
This follows the overview from earlier in the paper, with one main change.
In the earlier overview, 
we described a computation that used only the the documents containing
a single bursty word $w$.
This, however, leads to status gradients (as functions of time) that
are quite noisy.
Instead, we compute a single, smoother aggregate status gradient
over all the bursty words in the dataset.

Essentially, we can do this simply by merging all the time-stamped
documents containing any of the bursty words, including each document
with a multiplicity corresponding to the number of bursty words it contains,
and shifting the time-stamp
on each instance of a document with bursty word $w$
to be relative to the start of the burst for $w$.
Specifically, each of the bursty words $w$ selected above has
a time $\beta_w$ at which its burst interval begins.
For each document containing $w$, produced at time $T$,
we define its {\em relative time} to be $T - \beta_w$ ---
i.e. time is shifted so that the start of the burst is at time $0$.
(Time is measured in integer numbers of weeks for all of our datasets
except DBLP and Arxiv, where it is measured in integer numbers of years and months, respectively.)

Now we take all the documents and we bucket them into groups
that all have the same relative time: for each document produced
at time $T$ containing a bursty word $w$, we place it in the bucket 
associated with its relative time $T - \beta_w$.\footnote{If 
a document contains multiple bursty words, we place it in multiple buckets.
Also, to reduce noise,
in a post-processing step we combine adjacent buckets
if they both have fewer than a threshold number of documents $\theta$,
and we continue this combining process iteratively from earlier to later buckets until all buckets
have at least $\theta$ documents.  In our analysis we use $\theta = 1500$.}
From here, the computation proceeds as in the overview earlier 
in the paper: for each relative time $t$, we consider the median
activity level $g(t)$ of all documents in the bucket associated with $t$.
This function $g(t)$ plays the role of the single-word function
$g_w(t)$ from the overview, and the computation continues from there.

\xhdr{Final and Current Activity Levels} 
The computation of the status gradient involves the activity levels
of users, and there are two natural ways to define this quantity,
each leading to qualitatively different sets of questions.
The first is the {\em final activity level}:
defining each user's activity level to be 
the lifetime number of documents they produced. Under this interpretation,
an author will have the same activity whenever we see them in the data,
since it corresponds to their cumulative activity.
This was implicitly the notion of activity level that 
was used to describe the status gradient computation earlier.

An alternate, also meaningful, way to define an author's activity
level is to define it instantaneously at any time $t$ to be
the number of documents the author has produced up to time $t$.
This reflects the author's involvement with the community at
the time he or she produced the document, but it does not show
his or her eventual activity in the community.

\begin{figure*}[htp]
\centering
\includegraphics[width=\textwidth]{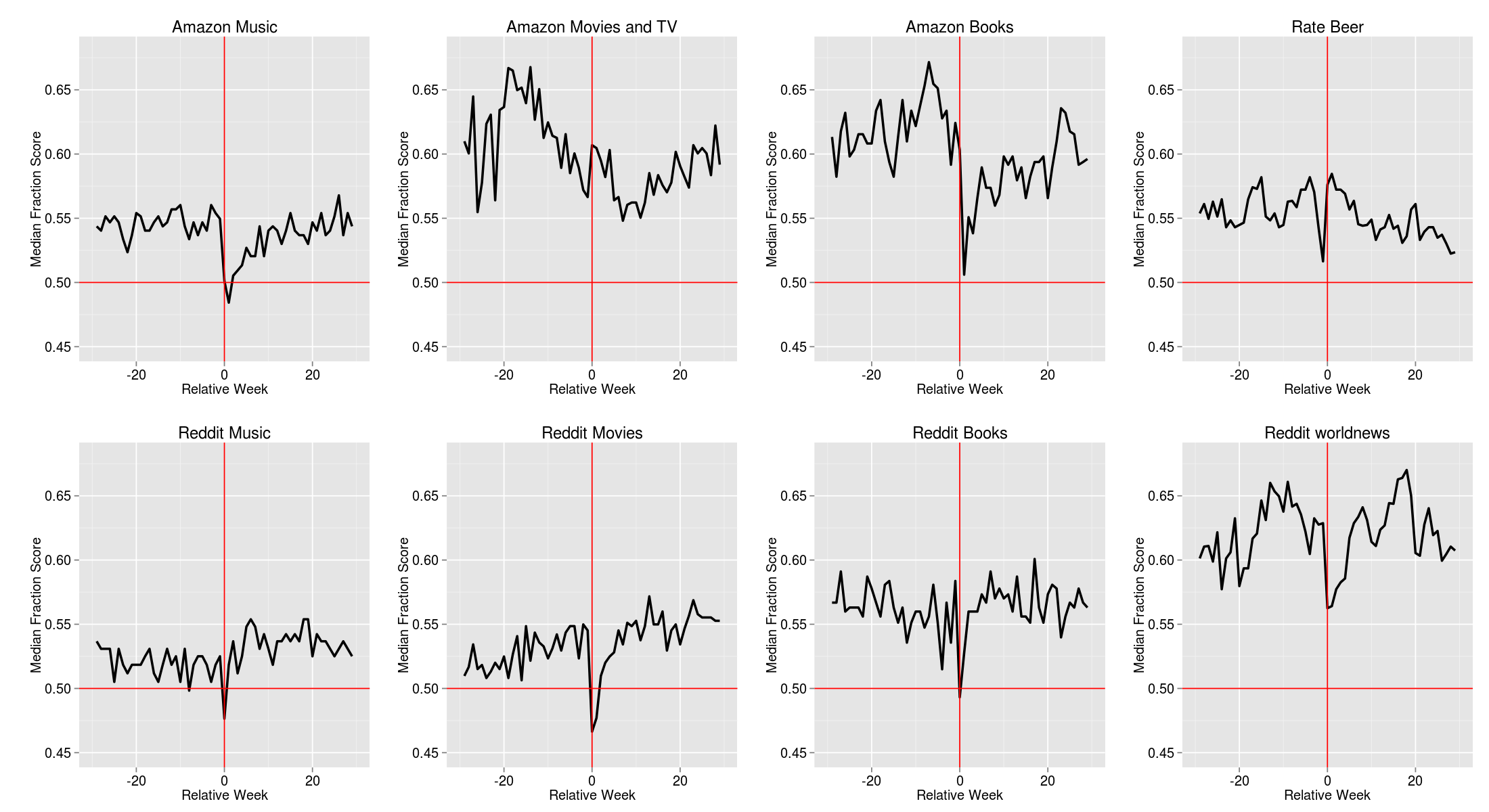}
\caption{
The status gradients for datasets from Amazon, Reddit, and an on-line beer
community, based on the final activity level of users and
a ranked set of 500 bursty words for each dataset.
}
\label{figure:FinalExperience}
\end{figure*}


Performing the analysis in terms of the final activity level is
straightforward.
For the analysis in terms of the current activity level, we 
need to be careful
about a subtlety.  If we directly adapt the method described so far,
we run into the problem that users' current activity levels are
increasing with time, resulting in status gradient 
plots that increase monotonically for a superficial reason.
To handle this issue, we compare
documents containing bursty words with documents which were
written at approximately the same time. 
Document $d$ written at time $t_d$
that has a bursty word will be compared to documents written in
the same week as $d$. 
We say that the {\em fractional rank} of document $d$ is the
fraction of documents written in the same week $t_d$ whose authors
have a smaller current activity level than the author of $d$.
Note that the
fractional rank is independent of the trending word; it depends only
on the week. Now that each
document has a score that eliminates the underlying monotone increase,
we can go back to the relative time domain and use the same
method that we employed for the final activity level, but using 
the fractional rank instead of the final activity level. 
Note that in this computation we thus have an extra level of indirection 
--- once for finding the fractional rank and a second time 
for computing the status gradient function.

As it turns out, the analyses using final and current activity
levels give very similar results; 
due to this similarity, we focus here on the computation and results
for the final activity level.

\xhdr{Bigrams}
Thus far we have performed all the analysis using trends
that consist of single words (unigrams).
But we can perform a strictly analogous computation in which
the trends are comprised of bursty two-word sequences (bigrams),
after stop-word removal.
Essentially all aspects of the computation remain the same.
The top 5 bigrams that the algorithm finds
are shown in Table \ref{TopWords}.
The results for bigrams in all datasets are very similar
to those for unigrams, and so in what follows we focus
on the results for unigrams.

\xhdr{DBLP and Arxiv} 
Compared to other datasets that we use in this study, DBLP and Arxiv have a
different structure in ways that are useful to highlight. 
We will point out two main differences.

First, documents on DBLP/Arxiv  generally only arrive in yearly/monthly increments,
rather than daily or weekly increments in the other datasets, and so we
perform our analyses by placing documents into buckets corresponding
to years/month rather than weeks.
In our heuristics for burst detection on DBLP, we require a minimum burst length
of 3 years (in place of the previous requirement of 8 weeks).
We found it was not necessary to use any additional minimum-length filters.

The second and more dramatic structural difference from the 
other datasets is that a given document will generally have 
multiple authors.
To deal with this issue, we adopt the following simple
approach: We define the current and final activity level of a document
as the highest current and final activity level, respectively,
among all its authors.\footnote{The results for taking the median experience instead of the maximum for each paper leads to similar results.}
Note, however, that a document still contributes to the activity level of
all its authors.

We observe that the bursty words identified 
for these datasets appear in at least 70
documents each instead of the minimum 200 we saw for the other datasets. 
We scaled down other parameters accordingly, and did not compute
bursty bigrams for DBLP and Arxiv.

\begin{figure}[htp]
\centering
\includegraphics[width=\columnwidth]{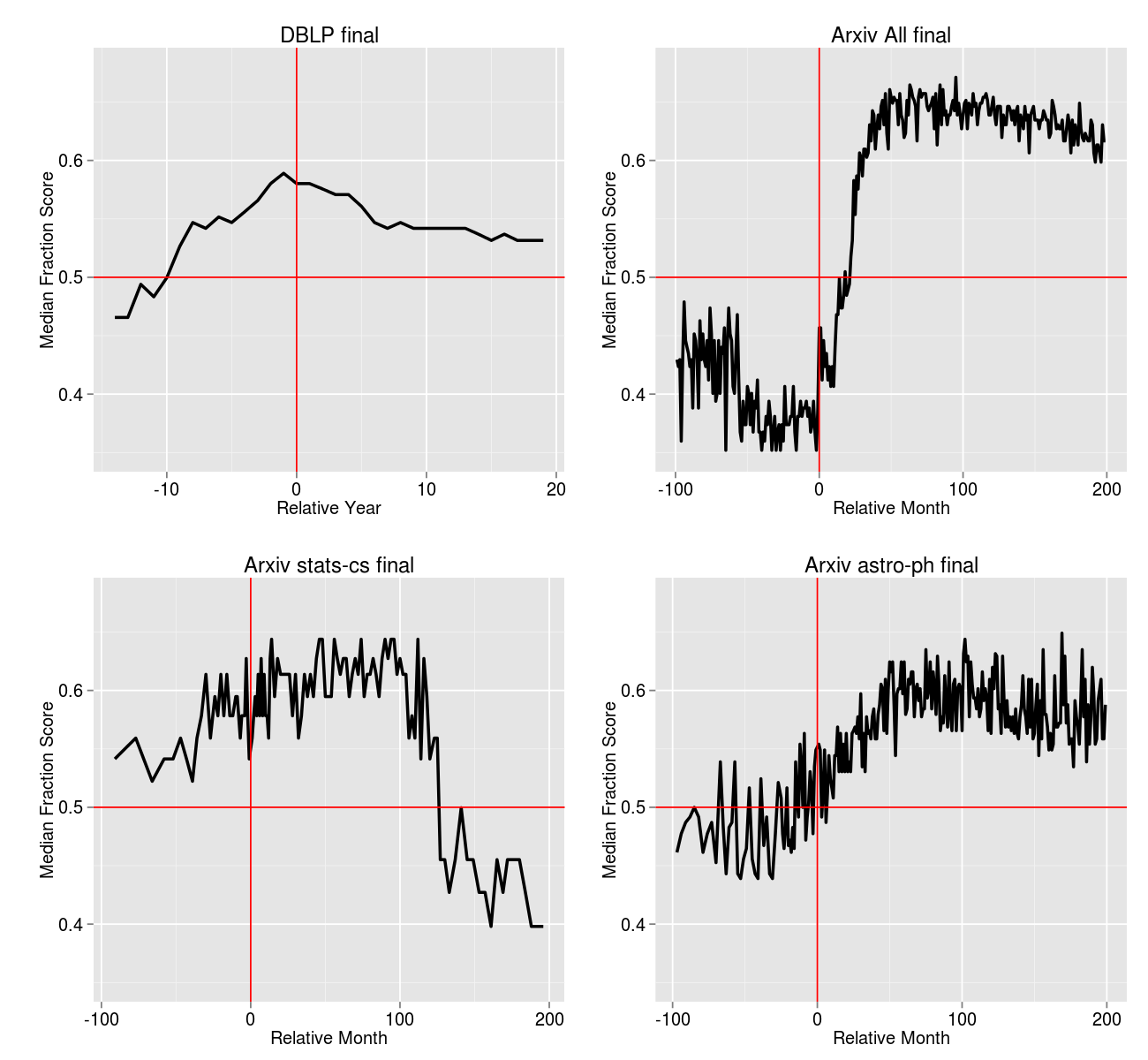}
\caption{
The status gradient for DBLP and Arxiv papers, as well as the 
stats-cs and astro-ph subsets of Arxiv, using final activity levels.
}
\label{figure:DBLP}
\end{figure}

\begin{figure*}[ht]
\centering
\includegraphics[width=\textwidth]{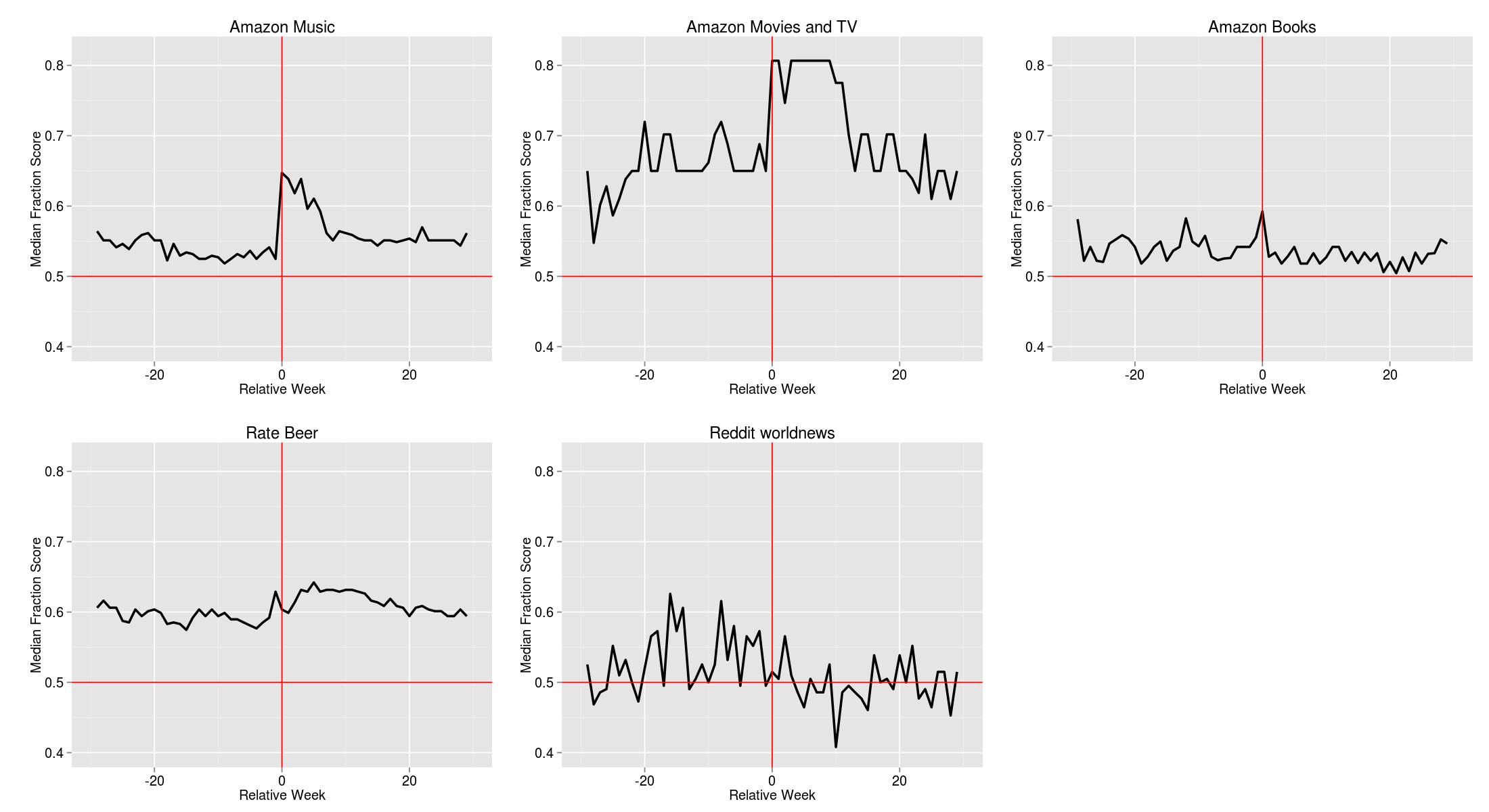}
\caption{Status gradients for producers --- brands on Amazon and
the beer community, and domains for Reddit World News.
As functions of time, these status gradients show strong contrasts
with the corresponding plots for the activity levels of 
users (consumers).
}
\label{figure:ProducerConsumer}
\end{figure*}

\section{Results}

Now that we have a method for computing status gradients, we 
combine the curves $f_w(t)$ over the top bursty words in each dataset,
as described above, 
aligning each bursty word so that time $0$ is the start of its burst, $\beta_w$.
In the underlying definition of the status gradient,
we focus here on the final activity level of users; the
results for current user activity are very similar.

\subsection{Dynamics of Activity Levels}

The panels of Figure \ref{figure:FinalExperience} show the aggregate status
gradient curves for the three Amazon categories, four of the
sub-reddits, and one of the beer communities.
(Results for the other sub-reddits and beer communities are similar.)

The plots in Figure \ref{figure:FinalExperience} exhibit two key
commonalities.

\begin{itemize}
\item 
First, they lie almost entirely above the line $y = 1/2$.
Recalling the definition of the status gradient, this means 
that high-activity individuals are using bursty words at a rate
{\em greater} than what their overall activity level would suggest.
That is, even relative to their already high level of
contribution to the site, the most active users are additionally
adopting the trending words.

\item 
However, there is an important transition in the curves right
at relative time $t = 0$, the point at which the burst begins.
For most of these communities there is a sharp drop, indicating
that the aggregate final activity level of users
engaging in the trend is abruptly reduced as the trend begins.
Intuitively, this points to an influx of lower-activity users
as the trend starts to become large.
This forms interesting parallels with related phenomena in cases
where users pursue content that has  become
popular \cite{aizen-batting,byers-groupon}.
\end{itemize}

This pair of properties --- overrepresentation of high-activity
users in trends (even relative to their general activity level);
and an influx of lower-activity users at the onset of the trend ---
are the two dominant dynamics that the status gradient reveals.
Relative to these two observations, we now identify a further
crucial property, the distinction between producers and consumers.

\subsection{Producers vs. Consumers}

We noted that the academic domains we study exhibit a considerably
different status gradient.  On DBLP (Figure \ref{figure:DBLP}),
the activity level of authors rises to a maximum very close
to relative time $t = 0$,
indicating an influx of high-activity users right at the start of a trend. 
Arxiv stats-cs shows the same effect, and 
the other Arxiv datasets show a time-shifted version of this pattern,
increasing through time $0$ and reaching a maximum shortly afterward.
(This time-shifting of Arxiv relative to DBLP may be connected
to the fact that Arxiv contains preprints while DBLP is a record
of published work, which may therefore have been in circulation 
for a longer time before the formal date of its appearance.)

This dramatic contrast to the status gradients in 
Figure \ref{figure:FinalExperience} highlights the fact that there is
no single ``obvious'' behavior at time $t = 0$, the start of the trend.
It is intuitive that low-activity users should rush in at the start of
a trend, as they do on Amazon, Reddit, and the beer communities;
but it is also intuitive that high-activity users should arrive 
to capitalize on the start of a trend, as they do on DBLP and Arxiv.
A natural question is therefore whether there is an underlying structural
contrast between the domains that might point to further analysis.

Here we explore the following contrast.  We can think of 
the users on Amazon, Reddit, and the beer communities as 
{\em consumers} of information: they are reviewing or commenting
on items (products on Amazon, generally links and news items on Reddit,
and beers on the beer communities) that are being produced by
entities outside the site.
DBLP and Arxiv are very different: its bibliographic data is tracking the activities of 
{\em producers} --- authors who produce papers for consumption 
by an audience.
Could this distinction between producers and consumers be relevant
to the different behaviors of the status gradients?

To explore this question, we look for analogues of producers
in the domains corresponding to Figure \ref{figure:FinalExperience}:
if the status gradient plots in that figure reflected populations of consumers, 
who are the corresponding producers in these domains?
We start with Amazon; for each review, there is not just an {\em author}
for the review (representing the consumer side) but also the 
{\em brand} of the product being reviewed (serving as a marker for
the producer side).  
We can define activity levels for brands just as we did
for users, based on the total number of reviews this brand is
associated with, and then use this in the Amazon data
to compute status gradients for brands rather than for users.

The contrasts with the user plots are striking, as shown in 
Figure \ref{figure:ProducerConsumer},
and consistent with
what we saw on DBLP and Arxiv: the status gradients for producers on Amazon
go up at time $t = 0$, and for two of the three categories 
(Music and Movies/TV), the increase at $t = 0$ is dramatic.
This suggests an interesting producer-consumer dynamic in bursts
on Amazon, characterized by a simultaneous influx of high-activity
brands and low-activity users at the onset of the burst: 
the two populations move inversely at the trend begins.
Intuitively, the onset of a burst is
characterized by producers of rising activity level moving in to provide content
to consumers of falling activity level.

We can look for analogues of producers in the other two domains
from Figure \ref{figure:FinalExperience} as well.
For Rate Beer, each review is accompanied by the brand of the beer,
and computing status gradients for brands we find a mild increase
at $t = 0$ here too --- as on Amazon, contrasting sharply with the
drop at $t = 0$ for the user population.
For Reddit, it is unclear whether there is a notion of a ``producer''
as clean as brands in the other domains, but for Reddit World News,
where most posts consist of a shared link, we can consider the domain
of the link as a kind of producer of the information.
The status gradient for domains on Reddit World News is noisy
over time, but we see a generally flat curve at $t = 0$;
while it does not increase at the onset of the trend, it again
contrasts sharply with the drop at $t = 0$ in the user population.
\begin{figure}[!ht]
\centering
\includegraphics[width=0.4\textwidth]{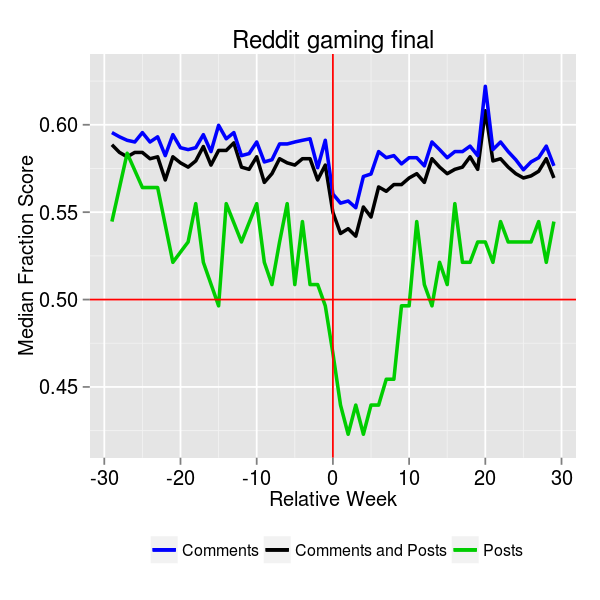}
\caption{A comparison between
the status gradients computed from posts,  comments, 
and the union of posts and comments on a large sub-reddit (gaming) .
}
\label{figure:postVSComment}
\end{figure}

\subsection*{Posters vs. Commenters}

As a more focused distinction, we can also look at contrasts
between different sub-populations of users on certain of the sites.
In particular, since the text we study on Reddit comes from 
threads that begin with a post and are followed by a sequence of
comments, we can look at the distinction between the status gradients
of posters and commenters.

We find (Figure \ref{figure:postVSComment})
that high-activity users are overrepresented 
more strongly in the bursts in comments than in posts;
this distinction is relatively minimal long before the burst,
but it widens as the onset of the burst approaches, and
the drop in the status gradient at $t = 0$ is much more strongly manifested
among the posters than the commenters.
This is consistent with a picture in which lower-activity
users initiate threads via posts, and higher-activity users
participate through comments, with this disparity becoming
strongest as the trend begins.

\subsection*{Life stages}
\begin{figure*}[ht	]
\centering
\includegraphics[width=\textwidth]{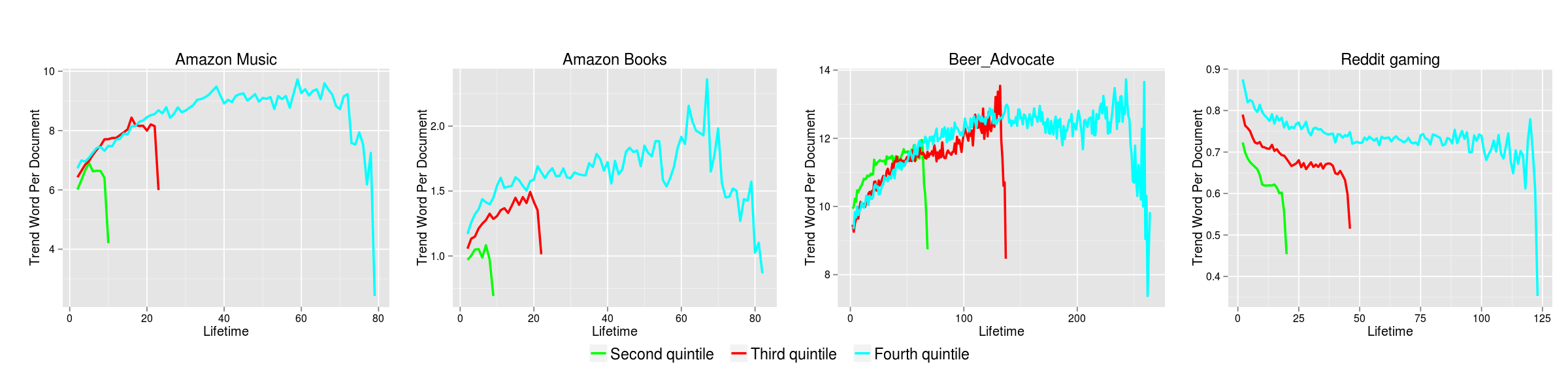}
\caption{The average number of bursty words used per document,
as a function of the author's life stage in the community.}
\label{figure:Cristian}
\end{figure*}

As a final point, we briefly consider a version of the dual question 
studied by Danescu-Niculescu-Mizil 
et al \shortcite{danescu-no-country-www13} ---
rather than tracking the life cycles of the words, as we have done so far,
we can look at the life cycles of the users and investigate how they
use bursty words over their life course on the site.
One reason why it is interesting to compare to this earlier work using
a similar methodology is that we are studying a related but fundamentally
different type of behavior from what they considered.
The word usage that they focused on can be viewed as
{\em lexical innovations}, or novelties,
in that they are words that had never been used 
before at all in the community.
Here, on the other hand, we are studying trending word usage through
the identification of bursts ---
the words in our analysis might have been used a non-zero number of
times prior to the start of the burst, but they grew dramatically
in size when the burst began, thus constituting trending growth.
It is not at all clear a priori that users' behavior with respect to
bursty words over their lifetime should be analogous to their
behavior with respect to novelties, but we can investigate this
by adapting the methodology from Danescu-Niculescu-Mizil et al 
\shortcite{danescu-no-country-www13}.

Here is how we set up the computation.
First, we remove
any authors (together with the documents they have written)
if their final activity
level is less than 10, since their life span is too short to analyze.
Then, we find four cut-off values that divide authors into 
{\em quintiles} --- five groups
based on their final activity level such that each group has produced a
fifth of the remaining documents. 
We focus on the middle three of these quintiles: three groups of
different final activity levels who have each collectively contributed
the same amount of content.

We then follow each author over a sequence of brief {\em life stages},
each corresponding to the production of five documents.
For each life stage and each quintile we find the average number of bursty
words per document they produce.

We find that the aggregate use of bursty words over
user life cycles can look different across different communities.
A representative sampling of the different kinds of patterns 
can be seen in Figure \ref{figure:Cristian}.
For many of the communities, we see the pattern noted by 
Danescu-Niculescu-Mizil et al, but adapted to bursty words 
instead of lexical innovations ---
the usage increases over the early part of a user's life cycle
but then decreases at the end.
For others, such as Reddit gaming shown in the figure, users
have the highest rate of adoption of bursty words at the beginning
of their life cycles, and it decreases steadily from there.
As with our earlier measures, these contrasts suggest the broader question 
of characterizing structural differences across sites through
the different life cycles of users and the trending words they adopt.

\section{Conclusions}

In this paper, we have a proposed a definition, 
the {\em status gradient}, and shown how it can be used
to characterize the adoption of a trend across
a social media community's user population.
In particular, it has allowed us to study the following
contrast, which has proven elusive in earlier work:
are trends in social media 
primarily picked up by a small number of the most active
members of a community, or by a large mass of less central
members who collectively account for a comparable amount of activity?
Our goal has been to develop a clean, intuitive
computational formulation of this question, in a manner that
makes it possible to compare results across multiple datasets.
We find recurring patterns, including a tendency for the most
active users to be even further overrepresented in trends,
and a contrast between the underlying dynamics for consumers versus
producers of information.

Because this work proposes an approach that is suitable
in many contexts, it also suggests a wide range of directions
for further work.
In particular, we have studied how the activity level of users
participating in a trend changes over time, but there
are many parameters of the trend that vary as time unfolds, 
and it would be interesting to track several of these at once
and try to identify relationships across them.
It would also be interesting to try incorporating the notion
of the status gradient
into formulations for the problem of starting or influencing a cascade,
building on theoretical work on this topic
\cite{domingos-kdd01,kempe-kdd03}.

\section{Acknowledgments}
We thank Cristian  Danescu-Niculescu-Mizil for valuable discussions,  Paul Ginsparg and Julian McAuley for their generous help with the Arxiv and 
Amazon dataset respectively, and Jack Hessel and Chenhao Tan 
for the Reddit dataset. This research was supported in part by a Simons Investigator Award, an ARO MURI grant, a Google Research Grant, and a Facebook Faculty Research Grant.
\bibliographystyle{aaai}
\bibliography{refs}   
\end{document}